\documentstyle[11pt]{article}
\makeatletter
\def\flE{\begin{picture}(0,0)
   \put( 0.25,    0){\vector( 1, 0){0.50}}
   \@ifstar{\@flE}{\@@flE}}
\def\@flE  #1{\put( 0.5 ,-0.03){\makebox(0,0)[ t]{$#1$}}\end{picture}}
\def\@@flE #1{\put( 0.5 , 0.03){\makebox(0,0)[ b]{$#1$}}\end{picture}}
\def\flNE{\begin{picture}(0,0)
   \put( 0.18, 0.18){\vector( 1, 1){0.64}}
   \@ifstar{\@flNE}{\@@flNE}}
\def\@flNE #1{\put( 0.52, 0.48){\makebox(0,0)[tl]{$#1$}}\end{picture}}
\def\@@flNE#1{\put( 0.48, 0.52){\makebox(0,0)[br]{$#1$}}\end{picture}}
\def\flN{\begin{picture}(0,0)
   \put(    0, 0.20){\vector( 0, 1){0.60}}
   \@ifstar{\@flN}{\@@flN}}
\def\@flN  #1{\put( 0.03, 0.5 ){\makebox(0,0)[ l]{$#1$}}\end{picture}}
\def\@@flN #1{\put(-0.03, 0.5 ){\makebox(0,0)[ r]{$#1$}}\end{picture}}
\def\flNW{\begin{picture}(0,0)
   \put(-0.18, 0.18){\vector(-1, 1){0.64}}
   \@ifstar{\@flNW}{\@@flNW}}
\def\@flNW #1{\put(-0.48, 0.52){\makebox(0,0)[bl]{$#1$}}\end{picture}}
\def\@@flNW#1{\put(-0.52, 0.48){\makebox(0,0)[tr]{$#1$}}\end{picture}}
\def\flW{\begin{picture}(0,0)
   \put(-0.25,    0){\vector(-1, 0){0.50}}
   \@ifstar{\@flW}{\@@flW}}
\def\@flW  #1{\put(-0.5 , 0.03){\makebox(0,0)[ b]{$#1$}}\end{picture}}
\def\@@flW #1{\put(-0.5 ,-0.03){\makebox(0,0)[ t]{$#1$}}\end{picture}}
\def\flSW{\begin{picture}(0,0)
   \put(-0.18,-0.18){\vector(-1,-1){0.64}}
   \@ifstar{\@flSW}{\@@flSW}}
\def\@flSW #1{\put(-0.52,-0.48){\makebox(0,0)[br]{$#1$}}\end{picture}}
\def\@@flSW#1{\put(-0.48,-0.52){\makebox(0,0)[tl]{$#1$}}\end{picture}}
\def\flS{\begin{picture}(0,0)
   \put(    0,-0.2 ){\vector( 0,-1){0.60}}
   \@ifstar{\@flS}{\@@flS}}
\def\@flS  #1{\put(-0.03,-0.5 ){\makebox(0,0)[ r]{$#1$}}\end{picture}}
\def\@@flS #1{\put( 0.03,-0.5 ){\makebox(0,0)[ l]{$#1$}}\end{picture}}
\def\flSE{\begin{picture}(0,0)
   \put( 0.18,-0.18){\vector( 1,-1){0.64}}
   \@ifstar{\@flSE}{\@@flSE}}
\def\@flSE #1{\put( 0.48,-0.52){\makebox(0,0)[tr]{$#1$}}\end{picture}}
\def\@@flSE#1{\put( 0.52,-0.48){\makebox(0,0)[bl]{$#1$}}\end{picture}}
\def\capsa(#1,#2)#3{\put(#1,#2){\makebox(0,0){$#3$}}}
\def\diagr{\@ifnextchar [{\@diagr}{\@diagr[15ex]}}
\def\@diagr[#1](#2,#3){\begingroup
   \setlength{\unitlength}{#1}
   \begin{picture}(#2,#3)}
\def\enddiagr{\end{picture}
   \endgroup}
\def\indiag{\@ifnextchar [{\@indiag}{\@indiag[15ex]}}
\def\@indiag[#1](#2,#3){\begingroup
   \setlength{\unitlength}{#1}
   \medskip
   \begin{center}
   \begin{picture}(#2,#3)}
\def\exdiag{\end{picture}
   \end{center}
   \medskip
   \endgroup}
\def\fflE{\begin{picture}(0,0)
   \put( 0.25,    0){\vector( 1, 0){1.50}}
   \@ifstar{\@fflE}{\@@fflE}}
\def\@fflE  #1{\put( 1   ,-0.03){\makebox(0,0)[ t]{$#1$}}\end{picture}}
\def\@@fflE #1{\put( 1   , 0.03){\makebox(0,0)[ b]{$#1$}}\end{picture}}
\def\qed{\ifvmode\removelastskip\fi
{\unskip\nobreak\hfil\penalty50\hbox{}\nobreak\hfil
\hbox{\vrule height1.2ex width1.2ex}\parfillskip=0pt
\finalhyphendemerits=0 \par\smallskip}}

\def\dif{{\rm d}}
\def\deriv{\@ifnextchar[{\@deriv}{\@deriv[]}}
   \def\@deriv[#1]#2#3{\mathchoice%
{{\dif^{#1}#2\over\dif{#3}^{#1}}}{{\dif^{#1}#2/\dif{#3}^{#1}}}%
{{\dif^{#1}#2\over\dif{#3}^{#1}}}{{\dif^{#1}#2/\dif{#3}^{#1}}}}
\def\derpar#1#2{\mathchoice%
{{\partial#1\over\partial#2}}{{\partial#1/\partial#2}}%
{{\partial#1\over\partial#2}}{{\partial#1/\partial#2}}}
\def\dderpar#1#2#3{\mathchoice%
{{\partial^2 #1\over\partial #2\,\partial #3}}%
{{\partial^2 #1/\partial #2\,\partial #3}}%
{{\partial^2 #1\over\partial #2\,\partial #3}}%
{{\partial^2 #1/\partial #2\,\partial #3}}}

\def\secteqno{\@addtoreset{equation}{section}%
\def\theequation{\thesection.\arabic{equation}}}
\newcounter{subequation}
\def\thesubequation{\alph{subequation}}
\def\sneqnarray{\stepcounter{equation}\let\@currentlabel=\theequation
\setcounter{subequation}{1}
\def\@eqnnum{{\rm (\theequation.\thesubequation)}}
\global\@eqcnt\z@\tabskip\@centering\let\\=\@eqncr\let\@@eqncr=\@@sneqncr
$$\halign to \displaywidth\bgroup\@eqnsel\hskip\@centering
 $\displaystyle\tabskip\z@{##}$&\global\@eqcnt\@ne
 \hskip 2\arraycolsep \hfil${##}$\hfil
 &\global\@eqcnt\tw@ \hskip 2\arraycolsep $\displaystyle\tabskip\z@{##}$\hfil
  \tabskip\@centering&\llap{##}\tabskip\z@\cr}
\def\endsneqnarray{\@@sneqncr\egroup $$\global\@ignoretrue}
\def\@@sneqncr{\let\@tempa\relax
   \ifcase\@eqcnt \def\@tempa{& & &}\or \def\@tempa{& &}
   \else \def\@tempa{&}\fi
     \@tempa \if@eqnsw\@eqnnum\stepcounter{subequation}\fi
     \global\@eqnswtrue\global\@eqcnt\z@\cr}
\makeatother
\def\artit#1{``#1'',}
\def\ben{\begin{enumerate}}
\def\een{\end{enumerate}}
\def\beq{\begin{equation}}
\def\eeq{\end{equation}}
\def\bea{\begin{eqnarray}}
\def\eea{\end{eqnarray}}
\def\beann{\begin{eqnarray*}}
\def\eeann{\end{eqnarray*}}
\def\beasn{\begin{sneqnarray}}
\def\eeasn{\end{sneqnarray}}

\newtheorem{prop}{Proposition}

\let\ds=\displaystyle
\def\buildord#1\over#2{\mathord{\mathop{\kern0pt #2}\limits^{#1}}}
\def\ir#1{{\buildord{\scriptscriptstyle\circ}\over #1}}
\def\map#1{\mathrel{\mathop{\to}\limits^{#1}}}
\def\mapping#1{\mathrel{\mathop{\longrightarrow}\limits^{#1}}}

\input amssym.def
\newsymbol\twoheadrightarrow 1310

\def\Id{{\rm Id}}
\def\pr{{\rm pr}}
\def\Real{{\bf R}}
\let\isom=\cong

\def\Ker{\mathop{\rm Ker}\nolimits}
\def\Hom{\mathop{\rm Hom}\nolimits}
\def\Lin{{\cal L}}
\def\transp#1{{}^{t}\kern-.15em\relax#1}
\def\Tens^#1{\buildord #1\over\otimes }

\def\Img{\mathop{\rm Im}\nolimits}

\def\Dif{{\rm D}}

\def\feble#1{\mathrel{\mathop{\simeq}\limits_{#1}}}
\def\scomp{\bullet}
\def\opers#1{(\!\!(#1)\!\!)}

\def\Tan{{\rm T}}
\def\Ver{{\rm V}}

\def\Newt{{\rm N}}

\def\vl{{\rm vl}}
\def\Lio{\Delta}

\def\calF{{\cal F}}

\def\calW{{\cal W}}

\def\bfp{{\bf p}}
\def\bfv{{\bf v}}
\def\sId{\mathord{\cal I\!\it d}}
\def\sDer{{\cal D}}
\def\sVer{{\cal V}}
\def\Der{{\rm D}}
\def\vv{^{\rm v}}

\def\ELform{\delta L}
\def\ELop{\widehat{\delta L}}
\def\Leg{\calF}
\def\Lleg{W}
\def\Hessop{\widehat{\cal\Lleg}}
\def\Ssehop{\widehat{\cal M}}
\def\En{E}
\def\gam{\gamma}
\def\Gam{\Gamma}
\def\Ups{\Upsilon}
\def\lam{\lambda}

\def\proof{\noindent{\it Proof}.\quad}

\title{Fibre derivatives: some applications to singular lagrangians}

\def\tabaddress#1{{\small\it\begin{tabular}[t]{c}#1\\[1.2ex]\end{tabular}}}
\def\UPCMAT{Departament de Matem\`atica Aplicada i Telem\`atica\\
   Universitat Polit\`ecnica de Catalunya\\
   Campus Nord UPC, edifici C3\\
   C. Jordi Girona, 1\\
   08034 Barcelona\\
   Catalonia, Spain}
\author{\sc
Xavier Gr\`acia
\\
\tabaddress{\UPCMAT}
\\[2mm]
\small e-mail: {\tt
xgracia@mat.upc.es}
}

\date{23 February 1999}

\secteqno

\begin{document}
\abovedisplayskip=6pt plus 3pt minus 1pt
\belowdisplayskip=\abovedisplayskip
\belowdisplayshortskip=4pt plus 3pt minus 1pt

\maketitle
\thispagestyle{empty}

\begin{abstract}

The fibre derivative of a bundle map is studied in detail.
In the particular case of a real function,
several constructions useful to study singular lagrangians
are presented.
Some applications are given;
in particular,
a geometric construction useful to solve
the Euler-Lagrange equation of a singular lagrangian.
A free particle in a curved space-time is studied as an example.

\bigskip
\parindent 0pt
\it

Keywords:
fibre derivative, bundle map, vertical bundle,
singular lagrangian, stabilisation algorithm.

MSC: 58C25, 58F05
\quad
PACS: 02.40.Vh, 03.20.+i

\end{abstract}

\section{Introduction}

During the last years
the tools of Differential Geometry
have been progressively applied to clarify the theory
of singular lagrangians of Theoretical Physics.
One of these tools is the fibre derivative.
In the general form that will be considered in this article,
the fibre derivative can be found in
\cite{GS-variations};
however, in the literature on lagrangian formalism
the fibre derivative has been mostly limited
to be the definition of the Legendre's transformation 
between the velocity space and the phase space,
whereas other geometric structures of the tangent and cotangent bundles
have played a more important role.
The purpose of this paper is twofold:
to give a thorough presentation of the fibre derivative
and its properties,
and to apply these results to some problems about singular lagrangians.

\medskip
To have a look at the question,
let us consider a fibre bundle
$P \to M$ and a real function $f \colon P \to \Real$.
If $\Ver(P)$ denotes the bundle of vertical tangent vectors of~$P$,
we shall see that $f$ defines a map
$\sVer f \colon P \to \Ver^*(P)$.
Indeed, 
if we write $f_x \colon P_x \to \Real$ for the restriction of
$f$ to the fibre~$P_x$,
then
$\sVer f(p_x) = \dif f_x(p_x)$,
which is an element of $\Tan_{p_x}^*(P_x) = \Ver_{p_x}^*(P)$.
Its local expression is
$\sVer f(x,p) = (x,p; \derpar fp)$. 

The construction of $\sVer f$ can not be extended
to higher-order fibre derivatives.
Let us consider again $f_x \colon P_x \to \Real$;
in general, it is not possible to define
something like the hessian of~$f_x$.
Of course, several objects may be constructed involving
(in coordinates) the second derivatives of $f_x$,
but not as a symmetric bilinear form on a certain space.
This is readily seen in coordinates:
if a change of coordinates $p \mapsto \bar p$ is performed,
with jacobian matrix $J = \derpar {p}{\bar p}$,
then the hessian matrix $\dderpar {f_x}pp$ becomes
$$
\dderpar {f_x}{\bar p^k}{\bar p^l} =
J^i_{\,k} \dderpar {f_x}{p^i}{p^j} J^j_{\,l} +
J^i_{\,k} \derpar {J^j_{\,l}}{p^i} \derpar {f_x}{p^j} .
$$
The annoying extra term disappears under two circumstances:
when $p$ is a critical point of~$f_x$,
and when the manifold $P_x$ has an appropriate extra structure
allowing for an atlas with constant jacobians.
For instance, when $P_x$ is an affine space.
Indeed,
it will be shown that higher-order fibre derivatives
can be defined for any bundle map $f \colon A \to B$,
where $A$ and $B$ are affine bundles over a manifold~$M$.
%

In the development of these concepts
we pay special attention to the fibre derivative of a fixed real function
$f \colon A \to \Real$.
Its fibre derivative is a map $\calF \colon A \to E^*$,
where $E$ is the vector bundle modelling the affine bundle~$A$, 
and we shall study several constructions concerning the fibre derivatives
of other functions defined either on $A$ or on~$E^*$.
Some of these constructions simplify other ones previously given in
\cite{CLR-geomstudy}
\cite{Gra-tesi}
\cite{GPR-higher}
to study several structures of first- and higher-order singular lagrangians.

\medskip
In the theory of singular lagrangians,
and more particularly in the study of the relations
between the lagrangian and the hamiltonian formalisms,
several interesting ideas have been presented in 
\cite{Kam-singular}
\cite{BGPR-equiv}
\cite{Pon-newrel}.
In particular, given a singular lagrangian~$L$,
the choice of a hamiltonian function $H$ and
a set of primary hamiltonian constraints $\phi_\mu$
allows to construct a local expression of 
a vector field $D_0$ in the lagrangian formalism
such that it satisfies the Euler-Lagrange equation on
the primary lagrangian constraint submanifold;
this vector field can be used as a departure point
for an algorithm to find
all the lagrangian constraints and the final dynamics.
It turns out that $D_0$ is coordinate-dependent;
however, in section~5
it will be shown that there is a geometric procedure
---based on the fibre derivative---
to construct a vector field that coincides with $D_0$
under a particular choice of coordinates.

\medskip
The paper is organised as follows.
Section~2 contains the definition and some properties
of the fibre derivative,
while section~3 is devoted to the particular case of a real function
and several special constructions.
In section~4 some relations between
lagrangian and hamiltonian formalisms are presented
and developed using the preceding constructions.
Section~5 addresses the problem of solving the Euler-Lagrange equation
for a vector field for a singular lagrangian.
Finally, these results are applied in section~6 
to the lagrangian of a free particle in a curved space-time.

Basic techniques about fibre bundles
are needed in what follows,
especially the pull-back of a bundle,
the tangent bundle of a bundle,
and particular properties of affine bundles.
They may be found in several books, as for instance
\cite{AMR-manif}
\cite{KMS-natural}
\cite{LR-higher}
\cite{Sau-jets}.

\section{Fibre derivatives}

As we have said in the introduction,
the concept of fibre derivative
can be found, for instance, in
\cite{GS-variations};
our presentation is more general and contains many details
to be used later.

\subsubsection*{The fibre tangent map}

Let $A \map{\pi} M$ and $B \map{\rho} M$ be fibre bundles over~$M$,
and $A \map{f} B$ a fibre bundle morphism.
Its tangent map $\Tan f \colon \Tan A \mapping{} \Tan B$
is a vector bundle morphism over~$f$.
Since
$\Tan(\pi) = \Tan(\rho) \circ \Tan(f)$,
if a vector $v_a \in \Tan_a(A)$ is vertical
({\it i.e.}, tangent to the fibres,
which amounts to saying that
$\Tan(\pi) \cdot v_a = 0$),
so it is $\Tan_a(f) \cdot v_a \in \Tan_{f(a)}(B)$.
Therefore $\Tan f$
restricts to a vector bundle morphism
between the bundles of vertical vectors,
$\Ver f \colon  \Ver A \to \Ver B$,
which can be called the {\it fibre tangent map} of~$f$.

Recall that this vector bundle morphism over $f$
is equivalent to a vector $A$-bundle morphism
---let us denote it by $\ir \Ver f$---
with values in the pullback of $\Ver B$,
$f^*(\Ver B) = A \times_f \Ver B$,
so we have:
\vskip -3mm
\indiag(2,0.3)
\capsa(0,0){\Ver A}
\capsa(-0.1,0){\flE{\ir \Ver f}}
\capsa(1,0){A \times_f \Ver B}
\capsa(1.1,0){\flE{}}
\capsa(2,0){\Ver B}
\put(0,0.2){\line(0,1){0.1}}
\put(0,0.3){\line(1,0){2}}
\put(2,0.3){\vector(0,-1){0.1}}
\put(1,0.33){\makebox(0,0)[b]{$\Ver f$}}
\exdiag
\noindent
Moreover, this vector $A$-bundle morphism
can also be identified with a section
---let us denote it by $\sVer f$---
of the vector $A$-bundle
$\Hom(\Ver A, A \times_f \Ver B) \to A$.
If the local expression of $f$ is
$(x^\mu,a^i) \mapsto (x^\mu, f^k(x,a))$,
then the local expression of the section $\sVer f$ is
$$
\sVer f (x^\mu,a^i) = \left( x^\mu, a^i, \derpar{f^k}{a^i}(x,a) \right) .
$$
\vskip -4mm
\vskip 0pt

\subsubsection*{The fibre derivative}

From now on our main concern is the following linear setting:
$f \colon A \to B$ is a fibre $M$-bundle map
between {\it affine bundles}\/
$\pi \colon A \to M$ and
$\rho \colon B \to M$
modelled respectively on vector bundles $E$ and~$F$.

Recall that the vertical bundle of the affine bundle $A \to M$
can be identified with the pull-back of $E$ to $A$ through
the vertical lift 
$\vl_A \colon A \times_M E \mapping{\cong} \Ver\! A$;
notice also for later use that a bundle map
$\xi \colon A \to E$ is canonically identified with a vertical 
vector field $\xi\vv$ on~$A$:
$\xi\vv(a_x) = \vl_A(a_x,\xi(a_x))$.

So, using also the isomorphism
$A \times_f (B \times_M F) \isom A \times_M F$,
we obtain a diagram
\indiag(2,1.3)

\put(0,1.2){\line(0,1){0.1}}
\put(0,1.3){\line(1,0){2}}
\put(2,1.3){\vector(0,-1){0.1}}
\put(1,1.33){\makebox(0,0)[b]{$\Ver f$}}

\capsa(-0.1,1){\Ver\! A}
\capsa(-0.1,1){\flE{\ir \Ver f}}
\capsa(1,1){A \times_f \Ver B}
\capsa(1.1,1){\flE{}}
\capsa(2.1,1){\Ver B}

\capsa(0,0){\flN{\isom}}
\capsa(2,0){\flN{\isom}}
\capsa(1,0.5){A \times_f (B \times_M F)}
\capsa(1,0.25){\llap{$\isom\,$}\uparrow}
\capsa(1,0.75){\llap{$\isom\,$}\uparrow}

\capsa(-0.2,0){A \times_M E}
\capsa(-0.1,0){\flE{\widehat\sDer f}}
\capsa(1,0){A \times_M F}
\capsa(1.1,0){\flE{}}
\capsa(2.2,0){B \times_M F}

\exdiag
\noindent
From this diagram we identify the vertical tangent map
with a vector $A$-bundle morphism $\widehat\sDer f$.
As before, this morphism can be identified with a section
$\sDer f$ of the vector bundle
$\Hom(A \!\times_M\! E,A \!\times_M\! F)$;
indeed we have
\indiag(3,1)

\capsa(-0.65,1){\Hom(\Ver\! A,A \!\times_f\! \Ver B)}
\capsa(0.15,1){\stackrel{\isom}{\to}}
\capsa(1.1,1){\Hom(A \!\times_M\! E,A \!\times_M\! F)}
\capsa(2,1){\stackrel{\isom}{\to}}
\capsa(2.68,1){A \!\times_M\! \Hom(E,F)}
\capsa(3.43,1){\stackrel{\pr_2}{\to}}
\capsa(3.98,1){\Hom(E,F)}

\capsa(0,0){\flN{\sVer f}}
\capsa(0.1,0){\flNE{\sDer f}}

\capsa(0,0){A}

\exdiag
\noindent
From $\sDer f$ and the maps in this diagram
we obtain an $M$-bundle map
\beq
\Der f \colon A \mapping{} \Hom(E,F) \isom F \otimes E^* ,
\eeq
which is called the {\it fibre derivative} of~$f$.
If the local expression of $f$ is
$(x^\mu,a^i) \mapsto (x^\mu,f^k(x,a))$,
then the local expression of $\Der f$ is
$$
\Der f(x^\mu,a^i) = \left( x^\mu, \derpar{f^k}{a^i}(x,a) \right) .
$$

Let us have a look at the fibres.
In the general case of an $M$-bundle morphism $f \colon A \to B$
between fibre bundles,
restriction to the fibres of $x \in M$ yields a manifold map
$f_x \colon A_x \to B_x$.
Its tangent map $\Tan_{a_x}(f_x)$ at $a_x \in A_x$ is identified with
a map $\Ver_{a_x}(A) \to \Ver_{f(a_x)}(B)$;
this is the fibre map $\Ver(f)_{a_x}$.
In the linear case
$A_x$ and $B_x$ are affine spaces,
and then the ordinary derivative of $f_x$
at a point $a_x \in A_x$ is a linear map
$\Dif f_x(a_x) \colon E_x \to F_x$
between the modelling vector spaces.
This map is the fibre derivative $\Der f(a_x)$ at $a_x$ as defined above.
See also
\cite{GS-variations}.

\subsubsection*{Higher order fibre derivatives}

Since $\Der f$ is also a bundle map between affine bundles,
the same procedure can be applied to compute its fibre derivative.
The canonical isomorphism
$\Hom(E,\Hom(E,F)) \isom \Lin^2(E;F)$
now yields the second fibre derivative,
the {\it fibre hessian},
which is a map
$$
\Der^2 f \colon A \mapping{} \Lin^2(E;F)
\isom \Hom(E \otimes E,F)
\isom F \otimes E^* \otimes E^* ,
$$
whose local expression is
$\ds
\Der^2 f(x^\mu,a^i) = \left( x^\mu, \dderpar{f^k}{a^i}{a^j}(x,a) \right) 
$.

More generally we obtain the $k$th order fibre derivative
\beq
\Der^k f \colon A \mapping{} \Lin^k(E;F)
\isom \Hom(\Tens^kE,F)
\isom F \otimes E^* \otimes \buildrel k\over\ldots \otimes E^* .
\eeq

The higher order fibre derivatives
can also be considered as sections of certain bundles,
and also as several vector $A$-bundle morphisms;
for instance, the hessian $\Der^2 f$ of $f\colon A \to B$
yields a section of
$A \times_M \Lin^2(E;F)$
and vector bundle maps like
$A \times_M (E \otimes E)  \to  A \times_M F$
and
$A \times_M E  \to  A \times_M \Hom(E,F)$.

\subsubsection*{Rules of derivation}

Some properties of the derivative apply to the fibre derivative.
For instance, if we have as before a bundle map
$f \colon A \to F$, and $\phi \colon A \to \Real$ is a function, then
\beq
\Der(f\phi) = (\Der f) \phi + f \otimes \Der \phi ,
\eeq
where now the last term is a map 
$A \to F \otimes E^* \isom \Hom(E,F)$.
Now,
if $\varphi\colon A \to F^*$ is another fibre bundle map,
then the contraction $\langle \varphi,f \rangle$ is a function on~$A$
whose fibre derivative is
\beq
\Der \langle \varphi,f \rangle =
\varphi \scomp \Der f +
\Der \varphi \scomp f ,
\eeq
where $\scomp$ denotes the composition between the images.

In the same way, if we consider three affine bundles bundles
and two bundle maps,
$A \map{f} B \map{g} C$, then
$\Der f \colon A \to \Hom(E,F)$
and we have
$\Der g \circ f \colon A \to \Hom(F,G)$,
and
\beq
\Der(g \circ f) = (\Der g \circ f) \scomp \Der f .
\eeq

\section{The fibre derivative of a real function}

Let us consider a fibre bundle $A \map{\pi} M$,
and a real function
$f \colon A \to \Real$.
This can be seen as a fibre bundle morphism
from $A$ to the trivial line bundle
$M \times \Real$.
Then the vertical tangent map is identified with a section
$\sVer f$ of $\Ver^* A$.
Indeed, $\sVer f$ is the composition of $\dif f$
with the canonical projection $\Tan^*A \to \Ver^*A$.

If $A$ is, as in the preceding section, an affine bundle
modelled on a vector bundle $E$,
then the fibre derivative of $f \colon A \to \Real$
is an $M$-bundle map
\beq
\calF = \Der f \colon A \mapping{} \Hom(E,M \times \Real) = E^* .
\eeq
In this section we shall study some properties of this map.
We begin with the following remark:
\begin{prop}
\label{prop-isom}
Let $\calF \colon A \to E^*$ be a fibre bundle map
---for instance the fibre derivative of $f \colon A \to \Real$.
There exists an isomorphism
$\Ver^*(A) \isom A \times_\calF \Ver(E^*)$
\end{prop}

\proof
The isomorphism is the composition of isomorphisms of vector $A$-bundles
\beq
\label{isom}
\ir b_\calF \colon
\Ver^*(A) \mapping{\transp{\vl_{A}}}
A \times_M E^* \mapping{\isom}
A \times_\calF (E^* \times_M E^*) \mapping{(\Id_A,\vl_{E^*})}
A \times_\calF \Ver(E^*) ,
\eeq
where the second one is due to the contravariance of the pull-back.
\qed

By projection to the second factor we obtain the vector bundle
morphism over~$\calF$
\beq
b_\calF 
\colon
\Ver^*(A) \mapping{} \Ver(E^*) ,
\label{b_F}
\eeq
which is an isomorphism at each fibre.

\subsubsection*{Relation between $\Tan(\calF)$ and the hessian}

Now we consider the fibre hessian $\Der^2 f$ of~$f$,
\beq
\Der^2 f = \Der \calF \colon A \mapping{} \Hom(E,E^*) ,
\eeq
whose associated vector $A$-bundle morphism is
$\widehat\sDer \calF \colon A \times_M E \to A \times_M E^*$.

To establish the relation between $\Der^2 f$ and $\Tan(\calF)$
it will be useful to consider the fibre hessian
as a morphism between the vertical vectors,
which is done through the vertical lift on~$A$.
So $\widehat\sDer \calF$ is identified 
with a symmetric vector $A$-bundle morphism
\beq
\widehat{\calW} \colon  \Ver(A) \mapping{} \Ver^*(A)
\eeq

On the other hand, the isomorphism
$\Ver^*(A) \isom A \times_\calF \Ver(E^*)$
identifies the section 
$\sVer \calF \colon A \to \Hom(\Ver\!A,A \times_\calF \Ver(E^*))$
---and therefore also $\sDer \calF$---
with a section
$$
{\calW} \colon A \to \Hom(\Ver(A),\Ver^*(A)) ,
$$
whose corresponding operator is indeed~$\widehat{\calW}$.

The following diagram shows the different morphisms involved:

\indiag(4.2,2)(-2.1,-1)

\capsa(0,1){A \!\times_\calF\! \Ver(E^*)}

\capsa(0,0.5){A \times_\calF (E^* \times_M E^*)}
\capsa(0,0.25){\llap{$\isom$}\uparrow}
\capsa(0,0.75){\llap{$\isom$}\uparrow}

\put(-1.80,0.15){\vector(2,1){1.4}}
\put(-1.11,0.52){\makebox(0,0)[br]{$\ir \Ver \calF$}}

\put(0.50,0.85){\vector(2,-1){1.4}}
\put(1.23,0.54){\makebox(0,0)[bl]{$\pr_2$}}

\capsa(-2.10,0){\Ver(A)}
\capsa(-2.20,0){\flE{\isom}}
\capsa(-1.15,0){A \!\times_M\! E}
\capsa(-1.10,0){\flE{\widehat\sDer \calF}}
\capsa( 0.00,0){A \!\times_M\! E^*}
\capsa( 0.15,0){\flE{(\calF,\Id_{E^*})}}
\capsa( 1.30,0){E^* \!\!\times_M\!\! E^*}
\capsa( 1.35,0){\flE{\isom}}
\capsa( 2.35,0){\Ver(E^*)}

\capsa(0,-1){\flN{\isom}}

\put(-1.80,-0.15){\vector(2,-1){1.4}}
\put(-1.11,-0.52){\makebox(0,0)[tr]{$\widehat \calW$}}

\put(0.50,-0.85){\vector(2,1){1.4}}
\put(1.23,-0.54){\makebox(0,0)[tl]{$b_\calF$}}

\capsa(0,-1){\Ver^*(A)}

\exdiag

Since $\calF$ is fibred over~$M$,
$
\Ker \Tan(\calF) \subset \Ker \Tan(\pi) = \Ver(A) 
$;
therefore the kernel of $\Tan(\calF)$ coincides with
the kernel of the fibre tangent map $\Ver(\calF)$.
Moreover, 
$$
\Ver(\calF) = b_\calF \circ \widehat{\calW}
$$
(where $b_\calF$ is the morphism defined by (\ref{b_F})),
therefore this kernel is also the kernel of $\widehat{\calW}$.
So we have proved:

\begin{prop}
With the notations above,
\beq
\Ker \Tan(\calF) = \Ker \widehat{\calW} .
\eeq
In particular,
$\calF$ is a local diffeomorphism at $a_x \in A$ iff
$\widehat{\calW}_{a_x}$ is a linear isomorphism.
\qed
\end{prop}

These statements follow also from the local expressions of the maps:
\beann
\calF : &&
(x,a) \mapsto \left(x,\hat \alpha(x,a) \right), \quad \hat\alpha_i = \derpar{f}{a^i} ,
\\
\Tan(\calF) : &&
(x,a;v,h) \mapsto \left(x,\hat \alpha(x,a); v, \dderpar fax v + \dderpar faa h \right) ,
\\
\Ver(\calF) : &&
(x,a;h) \mapsto \left(x,\hat \alpha(x,a); \dderpar faa h \right) ,
\\
\widehat\calW : &&
(x,a;h) \mapsto \left(x,a; \dderpar faa h \right) .
\eeann

\subsubsection*{The vector field $\Gam_h$}

Still let us consider the fibre derivative
$\calF \colon A \to E^*$ of~$f$.
We are going to derive several properties of a function
$h \colon E^* \to \Real$ and its fibre derivatives.

We use the notation
\beq
\gam_h = \Der h \circ \calF
\label{gam_h}
\eeq
for the composition
$
A \mapping{\calF} E^* \mapping{\Der h} E^{**} \isom E
$.
Recall that this map $\gam_h \colon A \to E$ yields in a canonical way,
through the vertical lift, 
a vertical vector field $\gam_h\vv$ on~$A$:
\beq
\Gam_h := \gam_h\vv = \vl_A \circ (\Id_A, \Der h \circ \calF)
\colon A \to A \times_M E \to \Ver\! A \subset \Tan A ,
\label{Gam_h}
\eeq
whose local expression is
$$
\Gam_h = \calF^* \left( \derpar{h}{\alpha_i} \right) \derpar{}{a^i}
$$
if the natural coordinates of $A$ and $E^*$ are respectively
$(x,a)$ and $(x,\alpha)$.

We also consider the composition
$$
A \mapping{\calF} E^* \mapping{\Der^2 h} \Hom(E^*,E^{**}) \isom \Hom(E^*,E) .
$$

Then, application of the chain rule yields
\beq
\Der (h \circ \calF) = \Der^2 f \scomp \gam_h ,
\label{D-h-F}
\eeq
\beq
\Der (\gam_h) = (\Der^2 h \circ \calF) \scomp \Der^2 f .
\label{D-gamh}
\eeq
(Here the composition is between the images of the maps
$\Der^2 f \colon A \to \Hom(E,E^*)$ and
$\Der^2 h \circ \calF \colon A \to \Hom(E^*,E)$.)
These properties will be used in the following section.

\smallskip

Notice from (\ref{D-h-F})
that if $h$ vanishes on the image $\calF(A) \subset E^*$
then $\Gam_h$ is in the kernel of $\widehat\calW$.
Then, we obtain the following result:

\begin{prop}
Let $f \colon A \to \Real$ with fibre derivative
$\calF \colon A \to E^*$.
Suppose that 
$\calF$ has connected fibres
and is a submersion onto a closed submanifold
---that is to say,
$f$ is almost regular in the terminology of
\cite{GN-preslag}.
Then the submanifold $\calF(A) \subset E^*$
can be locally defined by the vanishing of
a set of independent functions
$\phi_\mu \colon E^* \to \Real$,
and the vertical fields $\Gam_{\phi_\mu}$
constitute a frame for $\Ker \widehat\calW = \Ker \Tan(\calF)$.
\qed
\end{prop}

So we have recovered and generalised earlier results
from the theory of singular lagrangians
---see for instance
\cite{BGPR-equiv}.

\subsubsection*{The vector field along $\calF$ $\Ups^g$}

Now we present a construction dual to~$\Gam_h$.
Given a function $g \colon A \to \Real$, the map
\beq
\Ups^g = \vl_{E^*} \circ (\calF, \Der g)
\colon
A \to E^* \times_M E^* \to \Ver E^* \subset \Tan E^* ,
\label{Ups^g}
\eeq
is a vector field along the fibre derivative~$\calF$ of~$f$,
with local expression
$$
\Ups^g = \derpar{g}{a^i} \left( \derpar{}{\alpha_i} \circ \calF \right) .
$$
Notice also that
\beq
\Ups^g = b_\calF \circ (\Der g)\vv ,
\label{Ups-Der}
\eeq
where $b_\calF$ is defined by (\ref{b_F})
and
$(\Der g)\vv$ is the section of $\Ver^* \!A$
constructed from $\Der g$.
As differential operators, $\Gam_h$ and $\Ups^g$ are related by
$$
\Ups^g \cdot h = \Gam_h \cdot g .
$$

\subsubsection*{The Liouville's vector field}

Let $A$ be an affine bundle modelled on a vector bundle~$E$.
If $\eta \colon A \to E$ is a bundle map
with associated vertical field $Y = \eta\vv$ on~$A$,
and $g \colon A \to \Real$ is a function,
then it is easily seen that
\beq
Y \cdot g = \langle \Der g, \eta \rangle .
\label{Lie-ver}
\eeq

Recall that the Liouville's vector field of a vector bundle~$E$
is the vertical field $\Lio_E = \Id_E\vv$. 
If $g \colon E \to \Real$ is a function, 
then (\ref{Lie-ver}) yields
\beq
(\Lio_E \cdot g)(e_x) = \langle \Der g(e_x),e_x \rangle ,
\eeq
and then application of Leibniz's rule gives
\beq
\Der(\Lio_E \cdot g)(e_x) = \Der g(e_x) + \Der^2g(e_x) \cdot e_x .
\eeq

\section{Some structures relating lagrangian and hamiltonian
formalisms}

The basic concepts about singular lagrangian and hamiltonian formalisms
---Legendre's map, energy, hamiltonian function, hamiltonian constraints
\ldots---
are well-known and can be found in several papers,
for instance
\cite{Car-theory}.
First we shall recall some of these concepts.

\smallskip

Let us consider a first-order autonomous lagrangian
on a manifold~$Q$,
that is to say, a map
$L \colon \Tan Q \map{} \Real$.
Its fibre derivative (Legendre's map) and fibre hessian are maps
$$
\Tan Q \mapping{\Leg = \Der L} \Tan^*Q ,
$$
$$
\Tan Q \mapping{\Lleg = \Der \Leg} \Hom(\Tan Q,\Tan^*Q) .
$$
As said before,
$\Lleg$ can be identified with a vector bundle morphism
$\Hessop \colon \Ver(\Tan Q) \to \Ver^*(\Tan Q)$.
If this is an isomorphism
---equivalently, the Legendre's map is a local diffeomorphism---
the lagrangian $L$ is called {\it regular},
otherwise it is called {\it singular}.

\paragraph{Remark}

A $k$th order lagrangian is a function
$L \colon \Tan^k Q \to \Real$
\cite{LR-higher}
\cite{GPR-higher}.
Since the $k$th order tangent bundle
$\Tan^k Q$ is an affine bundle over $\Tan^{k-1}Q$
modelled on $\Tan^{k-1}Q \times_Q \Tan Q$,
the fibre derivative and hessian of $L$ can be studied
in a similar way, 
and some of the following developments could be extended to this case.
The case of a time-dependent lagrangian can also be dealed with in a
similar way.
Finally, some results could also be applied to field theory,
where the lagrangian density may be considered as a function 
on a certain affine bundle
\cite{GS-variations}.

\subsubsection*{The Euler-Lagrange form}

One of the basic objects of the variational problem associated to $L$
is the {\it Euler-Lagrange form}
---see for instance
\cite{CLM-higherNoether}
\beq
\ELform \colon \Tan^2 Q \to \Tan^*Q ;
\eeq
this is a 1-form along the second tangent bundle projection
$\tau_{2} \colon \Tan^2 Q \to Q$
with local expression
$$
\ds \ELform(q,v,a) =
\left( \derpar L{q^i} - \Dif_t \left( \derpar L{v^i} \right) \right) \dif q^i .
$$
A solution of the Euler-Lagrange equation is a path
$\gamma \colon I \to Q$
such that
$
\ELform \circ \ddot\gamma = 0 ,
$
where $\ddot\gamma \colon I \to \Tan^2 Q$
is the second time-derivative of~$\gamma$.

\medskip

It is often convenient to consider
second-order differential equations
as first-order equations on the tangent bundle
represented by vector fields satisfying the second-order condition.
Let us denote by $\Newt(Q) \subset \Tan(\Tan Q)$
the subset of tangent vectors satisfying the second-order condition;
this is an affine subbundle modelled on
the vector subbundle $\Ver(\Tan Q)$ of vertical tangent vectors.
There is a canonical immersion $\Tan^2 Q \to \Tan(\Tan Q)$
that identifies $\Tan^2 Q$ with $\Newt(\Tan Q)$.
Since $\Tan Q \times_Q \Tan Q$ is also identified with $\Ver(\Tan Q)$,
we can regard the Euler-Lagrange form as a map
\beq
\ELop \colon \Newt(\Tan Q) \to \Ver^*(\Tan Q) ,
\eeq
which is indeed an affine bundle map;
its associated vector bundle morphism is
\beq
\vec{\ELop} = - \Hessop .
\label{ELHess}
\eeq

Let us remark that there are other geometric expressions
of the Euler-Lagrange equation,
using the presymplectic form $\omega_L$ defined by~$L$
or the time-evolution operator~$K$
---see for instance
\cite{Car-theory}
\cite{GP-gener}.
Equations relating $\ELform$ with these objects can also be obtained.

\subsubsection*{Connection with the hamiltonian space}

Let $h \colon \Tan^*Q \map{} \Real$ be a function.
Recall the notation
$\gam_h = \Der h \circ \Leg \colon \Tan Q \to \Tan Q$,
and that this map is canonically identified with a
vertical vector field $\Gam_h$ on $\Tan Q$.

We assume that $L$ is an almost regular lagrangian
\cite{GN-preslag};
this is the most basic technical requirement
to develop a hamiltonian formulation from
a singular lagrangian~$L$.
Then the image of the Legendre's map is a submanifold
$P_0 \subset \Tan^*Q$,
the {\it primary hamiltonian constraint submanifold}.
Recall from Proposition~3 that,
if $\phi_\mu$ constitute an independent set of
{\it primary hamiltonian constraints},
then the vertical fields $\Gam_\mu = \Gam_{\phi_\mu}$
constitute a frame for $\Ker \Hessop$ and also for $\Ker \Tan(\Leg)$.

\smallskip

Recall that the {\it energy}\/ of~$L$ is defined by
$
\En = \Lio_{\Tan Q} \cdot L - L .
$
Due to the properties of the Liouville's vector field
at the end of the preceding section,
\beq
\En(v_q) = \langle \Der L(v_q),v_q \rangle - L(v_q) ,
\eeq
\beq
\Der \En(v_q) = \Lleg(v_q) \cdot v_q .
\label{Der-En}
\eeq

Finally, recall that a {\it hamiltonian}\/ 
is a function $H \colon \Tan^*Q \to \Real$
such that
$
\En = H \circ \Leg .
$
It exists since $L$ is almost regular,
and is unique on the primary hamiltonian constraint submanifold.

\subsubsection*{A resolution of the identity}

Given an almost regular lagrangian~$L$,
the choice of suitable data
yields a (local) resolution of the identity map of~$\Tan Q$ as follows.

\begin{prop}
\label{prop-resol}
Let $L$ be an almost regular lagrangian,
$\phi_\mu$
a set of independent primary hamiltonian constraints
and $H$ a hamiltonian function
(on an open set of $\Tan^*Q$).
Then there exist functions $\lam^\mu$
(defined on an open set of $\Tan Q$)
such that, locally,
\beq
\Id_{\Tan Q} = \gam_H + \sum_{\mu} \gam_\mu \, \lam^\mu .
\label{lam}
\eeq
Moreover,
\beq
\sId_{\Hom(\Tan Q,\Tan Q)} =
M \scomp \Lleg + \sum_\mu \gam_\mu \otimes \Der \lam^\mu ,
\label{IMW}
\eeq
where
\beq
M =
(\Der^2 H \circ \Leg) + \sum_\mu (\Der^2 \phi_\mu \circ \Leg) \, \lam^\mu .
\label{M}
\eeq
\end{prop}
(Notice that $\Lleg$ is a map $\Tan Q \to \Hom(\Tan Q,\Tan^*Q)$
and $M$ is a map $\Tan Q \to \Hom(\Tan^*Q,\Tan Q)$.)

\proof
Applying the chain rule (\ref{D-h-F}) to the definition of~$H$ yields
$$
\Der \En(v_q) = \Lleg(v_q) \cdot \gam_H(v_q) ,
$$
so using (\ref{Der-En}) we obtain
$$
\Lleg(v_q) \cdot (v_q - \gam_H(v_q)) = 0 ,
$$
and since the terms in parentheses are in $\Ker \Lleg(v_q)$,
there exist numbers $\lam^\mu(v_q)$ such that
$
v_q = \gam_H(v_q) + \sum_{\mu} \gam_\mu(v_q) \lam^\mu(v_q)
$,
which is (\ref{lam}).

Its fibre derivative is equation (\ref{IMW}), which follows
by applying equation (\ref{D-gamh})
and the Leibniz's rule to (\ref{lam}).
\qed

The above proposition can be given a slightly different form,
using the identification of bundle maps $\Tan Q \to \Tan Q$
with vertical vector fields:
equation (\ref{lam})
can be rewritten as
\beq
\Lio_{\Tan Q} = \Gam_H + \sum_{\mu} \Gam_\mu \, \lam^\mu .
\label{lam'}
\eeq
In the same way, equation (\ref{IMW}) can be expressed as
an endomorphism of $\Ver(\Tan Q)$:
\beq
\Id_{\Ver(\Tan Q)} =
 \Ssehop \circ \Hessop + \sum_\mu \Gam_\mu \otimes (\Der \lam^\mu)\vv ,
\label{IMW'}
\eeq
where
$\Ssehop \colon \Ver^*(\Tan Q) \to \Ver(\Tan Q)$
is the operator corresponding to the map~$M$,
and
$(\Der \lam^\mu)\vv$ is the section of $\Ver^*(\Tan Q)$
deduced from the map $\Der \lam^\mu \colon \Tan Q \to \Tan^*Q$.

\smallskip

Notice that
application of (\ref{IMW}) to $\gam_\nu$ yields
$\gam_\nu = \sum_\mu \gam_\mu \langle \Der \lam^\mu, \gam_\nu \rangle$;
then equation (\ref{Lie-ver}) shows that
\beq
\Gam_\nu \cdot \lam^\mu = \delta^{\mu}_{\,\nu} .
\eeq

Now let us write the local expressions of (\ref{IMW}) and (\ref{M}):
$$
\delta^i_{\,k} =
\sum_j M^{ij} W_{jk} + \sum_\mu
 \Leg^*\left(\derpar{\phi_\mu}{p_i}\right)  \derpar{\lam^\mu}{v^k} ,
$$
$$
M^{ij} = \Leg^* \left( \dderpar{H}{p_i}{p_j} \right) +
\sum_\mu \Leg^* \left( \dderpar{\phi_\mu}{p_i}{p_j} \right) \lam^\mu .
$$
These were deduced in
\cite{BGPR-equiv}
by derivating the local expression of
(\ref{lam}), which is
$$
v^i =
\Leg^* \left( \derpar{H}{p_i} \right) +
\sum_\mu \Leg^* \left( \derpar{\phi_\mu}{p_i} \right) \lam^\mu .
$$
See also
\cite{Kam-singular}.

\paragraph{Remark}

The manifold $\Tan^*(Q)$ has a
canonical symplectic 2-form.
Let $X_h$ be the hamiltonian vector field of a function~$h$.
Then
\beq
\Tan(\tau^*_Q) \circ X_h \circ \Leg \,\colon  \Tan Q \to \Tan Q .
\eeq
coincides with the map $\gam_h$ (\ref{gam_h});
this yields the construction of the vertical fields $\Gam_h$ 
that appeared in
\cite{CLR-geomstudy}.
A similar construction was given in
\cite{Gra-tesi}
for the maps $\Ups^g$ defined by (\ref{Ups^g}).

\section{The Euler-Lagrange equation for a vector field}

\subsubsection*{The equation of motion}

According to the preceding section, 
the Euler-Lagrange equation will be regarded as
as first-order equation on the tangent bundle.
If $X \colon \Tan Q \to \Newt(\Tan Q)$
is a second-order vector field,
its integral curves are solutions of the Euler-Lagrange equation
iff $\ELop \circ X = 0$
---recall that
$\ELop \colon \Newt(\Tan Q) \to \Ver^*(\Tan Q)$
is an affine bundle map.

If $L$ is singular then $\ELop$ is not an isomorphism,
therefore in general there is not
a vector field $X$ satisfying this equation everywhere.
It is more correct to consider the Euler-Lagrange equation
as an equation both for
a second-order vector field $X$ on $\Tan Q$ and
a submanifold $S \subset \Tan Q$ such that
\beq
\ELop \circ X \feble{S} 0 ,
\label{EL-X}
\eeq
supplemented with the condition that
\beq
\hbox{$X$ is tangent to $S$} .
\eeq
(As a matter of notation,
$\ds f \feble{S} g$
means that the maps $f$ and $g$ coincide on the subset~$S$.)

The {\it primary lagrangian constraint subset}\/ $V_1 \subset \Tan Q$
is the set of points $v \in \Tan Q$
such that there exists a second-order tangent vector $X_v$
such that $\ELop (X_v) = 0$;
it is assumed to be a closed submanifold.
Then (\ref{EL-X}) has solutions on~$V_1$;
for simplicity, any of these solutions will be called
a {\it primary dynamical field}.
Notice that these solutions may not be unique:
if $X_0$ is a fixed solution, then all the solutions are obtained by
adding any section of $\Ker \Hessop$ to~$X_0$.
On the other hand,
a primary dynamical field
is not necessarily tangent to~$V_1$;
we shall comment on this at the end of the section.
See
\cite{GP-gener}
for a careful discussion on these problems.

\subsubsection*{Construction of the primary lagrangian constraints}

The first step in order to solve the Euler-Lagrange equation
for a vector field is to determine
the primary lagrangian constraint submanifold.

\smallskip
If $A_0$ is an affine space modelled on $E_0$, $F_0$ is a vector space,
and $p_0 \in A_0$ is a fixed point,
the linear equation $f(p)=0$ (for $p \in A_0$)
is equivalent to
$\vec f(u) = -f(p_0)$ (for $u \in E_0$),
therefore its consistency condition is $f(p_0) \in \Img \vec f$.
If $(s_\mu)$ is a frame for $\Ker \transp{\vec f}$,
the consistency condition is equivalent to the vanishing of the numbers
$\langle s_\mu , f(p_0) \rangle$;
of course this does not depend on the point $p_0$ chosen.
\smallskip

We apply this remark to obtain
the consistency condition of the Euler-Lagrange equation
for a second-order vector field.
If $X_0$ is any fixed second-order vector field in $\Tan Q$,
the consistency condition is given by the vanishing of the functions
\beq
\chi_\mu =
\langle \ELop \circ X_0 , \Gam_\mu \rangle ,
\label{primlag}
\eeq
called the {\it primary lagrangian constraints}.
Their vanishing defines the primary lagrangian subset $V_1 \subset \Tan Q$,
assumed to be a submanifold.
Notice that these functions are not necessarily independent,
and indeed may vanish identically.

\paragraph{Remark}

The primary lagrangian constraints can be also constructed
without reference to a concrete second-order vector field.
Let
\beq
{\bar\chi}_\mu
= \langle \ELop , \Gam_\mu \circ \nu \rangle
= \langle \ELform , \gam_\mu \circ \tau_{21} \rangle \circ j_1^{-1} ,
\label{primlag'}
\eeq
where $\nu$ and $\tau_{21}$ are the projections of
$\Newt(\Tan Q)$ and $\Tan^2Q$ onto $\Tan Q$,
and $j_1$ is the diffeomorphism $\Tan^2Q \to \Newt(\Tan Q)$;
this defines functions ${\bar\chi}_\mu$
on the affine bundle $\Newt(\Tan Q)$,
and it is readily seen that they
are projectable to functions $\chi_\mu$ on $\Tan Q$,
since their fiber derivative vanishes.

\subsubsection*{Construction of primary fields}

Now we shall find a second-order vector field $X$ on $\Tan Q$ such that
$
\ELop \circ X \feble{V_1} 0
$.
If $X_0$ is a fixed second-order vector field,
giving $X$ amounts to giving a
vertical vector field $Y$ on $\Tan Q$ such that
$X = X_0 + Y$.
Due to (\ref{IMW'}),
the vertical field $Y$ can be written as
\beq
Y = \Ssehop \circ (\Hessop \circ Y) +
 \sum_\mu (Y \cdot \lam^\mu) \Gam_\mu .
\eeq
Moreover, by (\ref{ELHess})
$\ELop \circ (X_0 + Y) = \ELop \circ X_0 - \Hessop \circ Y$.
Therefore we have:
\beann
X
&=& X_0 + Y
\\
&=& X_0 + Y
 + \Ssehop \circ (\ELop \circ X_0 - \Hessop \circ Y)
 - \Ssehop \circ (\ELop \circ X)
\\
&=& X_0
 + \Ssehop \circ (\ELop \circ X_0)
 + \sum_\mu (Y \cdot \lam^\mu) \Gam_\mu
 - \Ssehop \circ (\ELop \circ X) .
\eeann
If $X$ is a primary field, then the last term is zero on~$V_1$,
and since the addition of vector fields of $\Ker \Lleg$
does not alter the set of primary fields,
we conclude that
$X_0 + \Ssehop \circ (\ELop \circ X_0)$ is also a primary field.
Let us give a more precise statement of this result:

\begin{prop}
Under the same hypotheses of Proposition \ref{prop-resol},
let $X_0$ be any second-order vector field on $\Tan Q$.
Then
\beq
\label{D0}
D_0 = X_0 + \Ssehop \circ (\ELop \circ X_0)
\eeq
is a primary field.
More precisely,
\beq
\ELop \circ D_0
= \sum_\mu \chi_\mu \, (\Der \lam^\mu)\vv
\label{D0-prim}
\feble{V^1} 0 ,
\eeq
where $\chi_\mu$ are the primary lagrangian constraints
(\ref{primlag}).
\end{prop}

\proof
Just apply (\ref{ELHess}) and the transpose of (\ref{IMW'}) to~$D_0$,
bearing in mind that $\Hessop$ and $\Ssehop$ are symmetric.
Then
\beann
\ELop \circ D_0
&=& \ELop \circ X_0 - \Hessop \circ \Ssehop \circ \ELop \circ X_0
\\
&=& \ELop \circ X_0 - \ELop \circ X_0
 + \sum_\mu \langle \ELop \circ X_0, \Gam_\mu \rangle \, (\Der \lam^\mu)\vv
\\
&=&
\sum_\mu \chi_\mu \, (\Der \lam^\mu)\vv .
\eeann
\vskip-\baselineskip\vskip 0pt\qed

In coordinates, if
$\ds X_0 = v \derpar{}{q} + A(q,v) \derpar{}{v}$,
then
$$
D_0 =
v \derpar{}{q} +
M \left(\derpar{L}{q} - \dderpar Lvq v\right) \derpar{}{v} +
\sum_\mu \left( \derpar{\lam^\mu}{v} A \right) \Gam_\mu .
$$
The coordinate-dependent choice of
$\ds X_0 = v^i \derpar{}{q^i}$ ($A=0$) yields a simpler expression for~$D_0$,
which is equation (4.19) in reference
\cite{BGPR-equiv}.
The resulting vector field was also used in
\cite{Pon-newrel}
to study some relations between lagrangian and hamiltonian formalisms.

\medskip
Once a primary field $D_0$ has been obtained,
all the primary fields are
\beq
D_u   \feble{V_1} D_0 + \sum_\mu u^\mu \Gamma_\mu ,
\label{Du}
\eeq
where $u^\mu$ are arbitrary functions.
The following step in the lagrangian stabilisation algorithm
is to study the tangency of $D_u$ to the submanifold~$V_1$;
this yields new constraints defining a submanifold $V_2 \subset V_1$
and determines some of the functions $u^\mu$ and so on.
The procedure follows the same lines as in
\cite{GNH-pres}
\cite{GP-gener},
and under some regularity assumptions finishes in a
final constraint submanifold $V_{\rm f} \subset \Tan Q$
and a family of second-order vector fields
$D^{\rm f} + \sum_{\mu_{\rm f}} u^{\mu_{\rm f}} \Gamma_{\mu_{\rm f}}$
tangent to $V_{\rm f}$ and
solution of the Euler-Lagrange equation (\ref{EL-X}).

\section{An example: a free particle
in a curved space-time}

In this simple example 
we will construct the geometric elements of the preceding sections,
and show how they can be used to solve the equation of motion.

Let $Q$ be a $d$-dimensional manifold
endowed with a metric tensor $g$ of signature $(1,d-1)$.
We write $v = \sqrt{g(\bfv,\bfv)}$,
which is a function defined on the open subset
$$
V = \{ \bfv \in \Tan Q \mid g(\bfv,\bfv)>0 \} \subset \Tan Q .
$$
The metric tensor defines an isomorphism
$\hat g \colon \Tan Q \to \Tan^*Q$.
We denote also by $g^*$ the 2-contravariant tensor deduced from~$g$.

The lagrangian of a free particle of mass $m$ in $Q$ is $L = mv$;
notice that it is defined only on~$V$.
Its fibre derivative (Legendre's map) and fibre hessian are:
\beq
V \mapping{\Leg} \Tan^*Q ,
\quad
\Leg(\bfv) = \frac{m}{v} \hat g(\bfv) ,
\eeq
\beq
V \mapping{\Lleg} \Hom(\Tan Q,\Tan^*Q) \isom \Tan^*Q \otimes \Tan^*Q ,
\quad
\Lleg(\bfv) = \frac{m}{v}\left(
    g \circ \tau_Q - \frac{1}{v^2}\hat g(\bfv)\otimes\hat g(\bfv) \right) .
\eeq
The lagrangian is singular, since
$\Ker \Lleg(\bfv)$ is spanned by~$\bfv$.

Using the vertical lift 
$\Tan Q \times_Q \Tan Q \mapping{\isom} \Ver(\Tan Q)$
we can extend the product of $g$ to vertical vectors.
So we obtain a section
$g^\Ver \colon \Tan Q \to 
\Hom(\Ver(\Tan Q),\Ver^*(\Tan Q)) \isom 
\Ver^*(\Tan Q) \otimes \Ver^*(\Tan Q)$,
and the corresponding vector bundle isomorphism
$\hat g^\Ver \colon \Ver(\Tan Q) \to \Ver^*(\Tan Q)$.
So we have, for instance,
$g^\Ver \opers{\Delta,\Delta} = v^2$,
since $\Delta$, the Liouville's vector field of~$\Tan Q$,
is the vertical lift of the identity.

In the same way,
we know that the hessian yields a section
${\cal W} \colon V \to 
\Hom(\Ver(\Tan Q),\Ver^*(\Tan Q)) \isom 
\Ver^*(\Tan Q) \otimes \Ver^*(\Tan Q)$,
and the corresponding operator
$\Hessop \colon \Ver(\Tan Q)|_V \to \Ver^*(\Tan Q)|_V$.
Then, if $Y$ is a vertical vector field on~$V$,
\beq
\Hessop\opers{Y} = 
\frac{m}{v} 
\left(
 \hat g^\Ver\opers{Y} - \frac{1}{v^2} 
 g^\Ver\opers{\Delta,Y} \hat g^\Ver\opers{\Delta}
\right) .
\eeq
 
Let $X$ be a second-order vector field.
The Euler-Lagrange operator 
$\ELop \colon \Newt(\Tan Q)|_V \to \Ver^*(\Tan Q)|_V$
is
\beq
\ELop\opers{X} = \Hessop\opers{S-X} ,
\label{partELop}
\eeq
where $S$ is the geodesic vector field of~$g$.
This can be obtained by computing the local expression
$\ds 
\ELform(q,v,a) =
\left(
 \derpar L{q^i} - \Dif_t \left( \derpar L{v^i} \right) 
\right) \dif q^i 
$,
and comparing with the local expression of the geodesic vector field,
which is
$$
S = 
v^\mu \derpar{}{q^\mu} -
\Gamma_{\alpha\beta}^\mu v^\alpha v^\beta \derpar{}{v^\mu} ,
$$
where $\Gamma_{\alpha\beta}^\mu$ are the Christoffel's symbols
of the Levi-Civita connection of~$g$.

Since $\Ker\Hessop$ is spanned by the Liouville's vector field,
from (\ref{partELop}) 
the Euler-Lagrange equation $\ELop\opers{X} = 0$
is easily solved:
\beq
X = S + \mu \Delta ,
\label{partXlagr}
\eeq
where $\mu \colon V \to \Real$ is any function.
The base integral curves of $X$ are 
the paths $\gamma$ in $Q$ satisfying
$\nabla_t \dot\gamma = (\mu \circ \dot\gamma) \dot\gamma$,
that is to say,
they are reparametrised geodesics.
The condition of being $\dot\gamma$ in $V$
is $g \opers{\dot\gamma,\dot\gamma} > 0$,
which means that $\gamma$ is a time-like curve. 

\medskip

Now let us see how the solution can be obtained from the hamiltonian
formalism and the procedure of the preceding section.

The image of the Legendre's map $\Leg$ 
is the submanifold $P_0 \subset \Tan^*Q$
defined by the vanishing of the hamiltonian constraint
\beq
\phi(\bfp) = \frac 12 ( g^*(\bfp,\bfp) - m^2 ) .
\eeq
Since the lagrangian $L$ is homogeneous of degree~1,
the lagrangian energy vanishes, and so does the hamiltonian $H$ on~$P_0$.

It is readily seen that
$\Der \phi(\bfp) = \hat g^{-1}(\bfp)$, therefore
\beq
\gam_\phi(\bfv) = \frac{m}{v} \bfv .
\eeq
From the identity (\ref{lam}),
$\Id_{\Tan Q} = \gam_H + \gam_\phi \, \lam^\phi$,
we obtain the function
\beq
\lam^\phi(\bfv) = \frac{v}{m} ,
\eeq
and thus
$\ds \Der \lam^\phi(\bfv) = \frac{1}{mv}\hat g(\bfv)$.
Finally, the map 
$M \colon V \to \Hom(\Tan^*Q,\Tan Q)$
defined by (\ref{M}) is
\beq
M(\bfv) = \frac{v}{m} \hat g^{-1} .
\eeq
One can then check that (\ref{IMW}) is satisfied.

Passing again to the vertical bundle,
from $\gam_\phi$ we obtain the vertical vector field
\beq
\Gam_\phi = \frac{m}{v} \Delta ,
\eeq
which spans the kernel of the fibre hessian
and therefore the kernel of $\Tan(\Leg)$.
We notice also that if $Y$ is a vertical vector field then
$\ds
\langle (\Der \lam^\phi)^\Ver , Y \rangle = 
\frac{1}{mv} g^\Ver\opers{\Delta,Y}$.
Finally, the operator
$\Ssehop \colon \Ver^*(\Tan Q)|_V \to \Ver(\Tan Q)|_V$
is given by
\beq
\Ssehop\opers{\Xi} = \frac{v}{m} (\hat g^\Ver)^{-1}\opers{\Xi} .
\eeq

Now we are ready to compute the composition
$\Ssehop \circ \Hessop \colon \Ver(\Tan Q)|_V \to \Ver(\Tan Q)|_V$:
$$
\Ssehop \circ \Hessop\opers{Y} = 
Y - \frac{g^\Ver\opers{\Delta,Y}\,\Delta}{v^2} ,
$$
that is to say, 
the orthogonal projection of $Y$
onto the subspace orthogonal to~$\Delta$.

Let us also show that there are no lagrangian constraints.
Indeed, from (\ref{primlag}), 
if $X_0$ is any second-order vector field
we obtain the primary lagrangian constraint as
$$
\chi = 
\langle \ELop \opers{X_0} , \Gam_\phi \rangle =
\langle \Hessop \opers{S-X_0} , \Gam_\phi \rangle =
\langle \Hessop \opers{\Gam_\phi} , S-X_0 \rangle = 0 ,
$$
where we have used that $\Hessop$ is symmetric.

Now, if $X_0$ is any second-order vector field,
application of (\ref{D0})
yields a lagrangian vector field
$$
D_0 
= X_0 + \Ssehop \circ (\ELop \circ X_0)
= X_0 + \Ssehop \circ \Hessop \opers{S-X_0}
= S - \frac{ g^\Ver \opers{\Delta,S-X_0} \Delta }{v^2} ,
$$
which is one of the solutions (\ref{partXlagr}) of the equation of motion.

\section*{Conclusions}

In this paper we have studied the fibre derivative
of a map between affine bundles.
This permits a careful study of several structures
constructed from the fibre derivative,
for instance its tangent map and the fibre hessian.

In the particular case of a singular lagrangian
we have applied these structures to obtain some geometric relations
between lagrangian and hamiltonian formalisms;
some of these relations were previously known in coordinates,
but not as geometric objects.
Since these developments work in any affine bundle,
they may be useful also for higher-order lagrangians and field theory.

\section*{Acknowledgements}

The author wishes to thank Profs.\
M.-C. Mu\~noz-Lecanda and J.M. Pons
for some useful comments.
He also acknowledges partial financial support
from CICYT TAP 97--0969--C03--01.



\end{document}